\documentclass[11pt]{article}
\usepackage[letterpaper,top=2cm,bottom=2cm,left=3cm,right=3cm,marginparwidth=1.75cm]{geometry}
\usepackage{graphicx}
\usepackage{authblk} 
\usepackage{tabularray}
\usepackage{threeparttable}
\usepackage{hyperref}
\usepackage{url}
\usepackage{color}
\usepackage{rotating}
\usepackage{CJKutf8}

\providecommand{\keywords}[1]
{
  \small	
  \textbf{Keywords:} #1
}

\title{The HS-CMU Dataset for Diagnosing Benign and Malignant Diseases through Hysteroscopy}
\author{Ruxue Han\textsuperscript{1}, Yuantao Xie\textsuperscript{2}, Kangze You\textsuperscript{2}, Lijun Cao\textsuperscript{2}\thanks{Corresponding author:caolj@jiyuanmedical.com}, Hua Li\textsuperscript{1}\thanks{Corresponding author:hual\_gyn@163.com}}

\affil{\textsuperscript{1}Department of Obstetrics and Gynecology, Beijing Chao-Yang Hospital, Capital Medical University, Beijing, China}
\affil{\textsuperscript{2}Jiangsu Jiyuan Medical Technology Co., Ltd}
\date{}
\begin{document}
\begin{CJK}{UTF8}{gbsn}

\maketitle

\begin{abstract}
Hysteroscopy enables direct visualization of morphological changes in the endometrium, serving as an important means for screening, diagnosing, and treating intrauterine lesions. Accurate identification of the benign or malignant nature of diseases is crucial. However, the complexity and variability of uterine morphology increase the difficulty of identification, leading to missed diagnoses and misdiagnoses, often requiring the expertise of experienced gynecologists and pathologists. Here, we provide the video and image dataset of hysteroscopic examinations conducted at Beijing Chaoyang Hospital, Capital Medical University (named the HS-CMU dataset), recording videos of 175 patients undergoing hysteroscopic surgery to explore the uterine cavity. These data were obtained using corresponding supporting software. From these videos, 3385 high-quality images from 8 categories were selected to form the HS-CMU dataset. These images were annotated by two experienced obstetricians and gynecologists using lableme software. We hope that this dataset can be used as an auxiliary tool for the diagnosis of intrauterine benign and malignant diseases.
\end{abstract}


\section{Value of the Data}
Hysteroscopy has entered a widely used era and has become an important diagnostic tool for benign and malignant diseases within the uterine cavity. Hysteroscopy provides direct visualization of the uterine cavity. Clinicians can assess and locate effective biopsy areas based on morphological changes in the endometrium under hysteroscopy, obtain pathological samples, and it is considered the standard for the diagnosis of benign and malignant diseases.
The screening and diagnosis of intrauterine diseases require experienced doctors, but the number of such doctors is limited, and clinical training of an experienced surgical doctor often requires a long accumulation of experience. Primary hospitals, especially in remote areas, suffer from a lack of pathologists, resulting in a severe deficiency in pathological diagnostic capabilities. With the rapid development of artificial intelligence in the field of medical imaging, assisted diagnostic algorithms based on deep learning of hysteroscopic images hold promise in addressing the issues.
This dataset includes hysteroscopic images and labels of various benign and malignant gynecological diseases. It can be used to develop software for the assisted diagnosis of benign and malignant intrauterine diseases based on hysteroscopic data, thereby achieving assisted diagnosis during hysteroscopic surgery.

\section{Introduction}
Abnormal uterine bleeding and infertility are common symptoms of suspected endometrial lesions in women. Hysteroscopy is a fiber light source endoscope, a minimally invasive technology that has a magnifying effect on the lesion and can obtain tissue for pathological diagnosis\cite{YEN20191480}. Because of its intuitiveness and accuracy, it has become the gold standard for the diagnosis of uterine cavity lesions\cite{Shen2023, Bosteels13}. If endometrial disease is not diagnosed promptly, treatment can be delayed, leading to disease progression. Endometrial polyps, submucosal fibroids, intrauterine adhesions, and endometrial hyperplasia are common benign uterine cavity diseases in gynecology. The incidence rate of endometrial cancer accounts for 4.5\% of all cancers globally, making it the sixth most common cancer among women worldwide\cite{cancer2020}. In recent years, its incidence rate has been increasing, and the age at diagnosis is becoming younger. Uterine cavity diseases are often associated with hormonal changes, injuries, infections, genetics, and other factors, and there are morphological differences among different uterine cavity diseases. The development of hysteroscopy has entered a widely used era. Hysteroscopy provides direct visualization of the uterine cavity, allowing clinicians to evaluate and locate effective biopsy areas based on morphological changes in the endometrium under hysteroscopy. For hysteroscopic examinations and surgeries, in addition to the improvement of surgical equipment, the experience of the operator, the ability to identify lesion tissues under the microscope, and the diagnostic techniques of the pathology department are crucial for the accurate assessment of uterine cavity diseases.

\section{Data Description}
\begin{table}[!ht]
\centering

\begin{tblr}{
  width = \linewidth,
  colspec = {Q[200]Q[800]},
  colspec = {Q[m]Q[m]},
  colspec = {Q[c]Q[l]},
  hlines,
  vlines,
  hline{1,9} = {-}{0.08em},
}

Subject  & Gynecology, Women’s Health \\
Specific subject area & Hysteroscopy has become an important diagnostic tool for benign and malignant diseases within the uterine cavity. \\
Type of data & Image and text \\
How the data were acquired & All images were taken using Karl Storz TC200 endoscopic camera systems with a resolution of 1920 × 1080 pixels and were stored in JPEG format. \\
Data format & JPEG \\
Description of data collection & A total of 175 videos from 175 patients were collected from Beijing Chaoyang Hospital, Capital Medical University from August 2023 to January 2024. All images were taken using Karl Storz TC200 endoscopic camera systems with a resolution of 1920 × 1080 pixels and were stored in MP4 format. Images are extracted from videos in 10 frames each. Finally, a total of 3385 images of different types of endometrial lesions were collected. \\ 
Data source location & Department of Obstetrics and Gynecology, Beijing Chao-Yang Hospital, Capital Medical University, Beijing, China. \\
Data accessibility & {Repository name: HS-CMU \\ Data identification number: https://openxlab.org.cn/datasets/jiyuanmedical/HS-CMU/}
\end{tblr}
\end{table}

\section{Dataset Statistics}
This dataset is publicly available at \url{https://openxlab.org.cn/datasets/jiyuanmedical/HS-CMU/}, which can be downloaded as a zip file. In the unzip file, There are two folders named as “image” and “label” are listed. The “image” folder contains the original images which are saved in the jpg format and named as “AB\_C.jpg” (C indicates which frame the image is in the video). And the “label” folder contains the labels of the corresponding images in the “image’’ folder and these labels are named as “AB\_C.txt”. In detail, “A” represents disease abbreviation for example SM (submucous myoma), EC (endometrial   cancer), EP (endometrial polyp), EPH (endometrial polypoid hyperplasia), EH (endometrial hyperplasia without atypical hyperplasia), IFB (intrauterine foreign body), CP (cervical polyp), AHE (atypical hyperplasia of endometrium). “B” represents case number. And “C” represents frame ID in video is file respectively. The numbers of eight types of images from each patient are listed in Table \ref{t1}. There are 3431 images, total including 4185 bounding boxes with label. In detail, 325 bounding boxes with the SM label, 352 bounding boxes with the EC label, 1330 bounding boxes with the EP label, 683 bounding boxes with the EPH label, 383 bounding boxes with the EH label, 466 bounding boxes with the IFB label, 550 bounding boxes with the CP label, 96 bounding boxes with the AHE label.

\begin{table}
\centering
\caption{\label{t1}Summary of the HS-CMU dataset. Different labels were assigned with different IDs (SM as 0, EC as 1, EP as 2, EPH as 3, EH as 4, IFB as 5, CP as 6, AHE as 7).}
\begin{threeparttable} 
\begin{tblr}{
  width = \linewidth,
  colspec = {Q[200]Q[80]Q[80]Q[80]Q[80]Q[80]Q[80]Q[80]Q[80]},
  cells = {c},
  cells = {m},
  hline{1,4} = {-}{0.08em},
  hline{2} = {-}{},
}
Label (ID)       & SM
(0) & EC
(1) & EP
(2) & EPH
(3) & EH
(4) & IFB
(5) & CP
(6) & AHE
(7) \\
Images num       & 287    & 268    & 1059   & 574     & 280    & 404     & 537    & 83      \\
Bounding box num & 325    & 352    & 1330   & 683     & 383    & 466     & 550    & 96      
\end{tblr}
\begin{tablenotes} 
\item[*] SM, submucous myoma; EC, endometrial cancer; EP, endometrial polyp; EPH, endometrial polypoid hyperplasia; EH, endometrial hyperplasia without atypical hyperplasia; IFB, intrauterine foreign body; CP, cervical polyp; AHE, atypical hyperplasia of endometrium.
\end{tablenotes}
\end{threeparttable}
\end{table}

\begin{figure}
    \centering
    \includegraphics{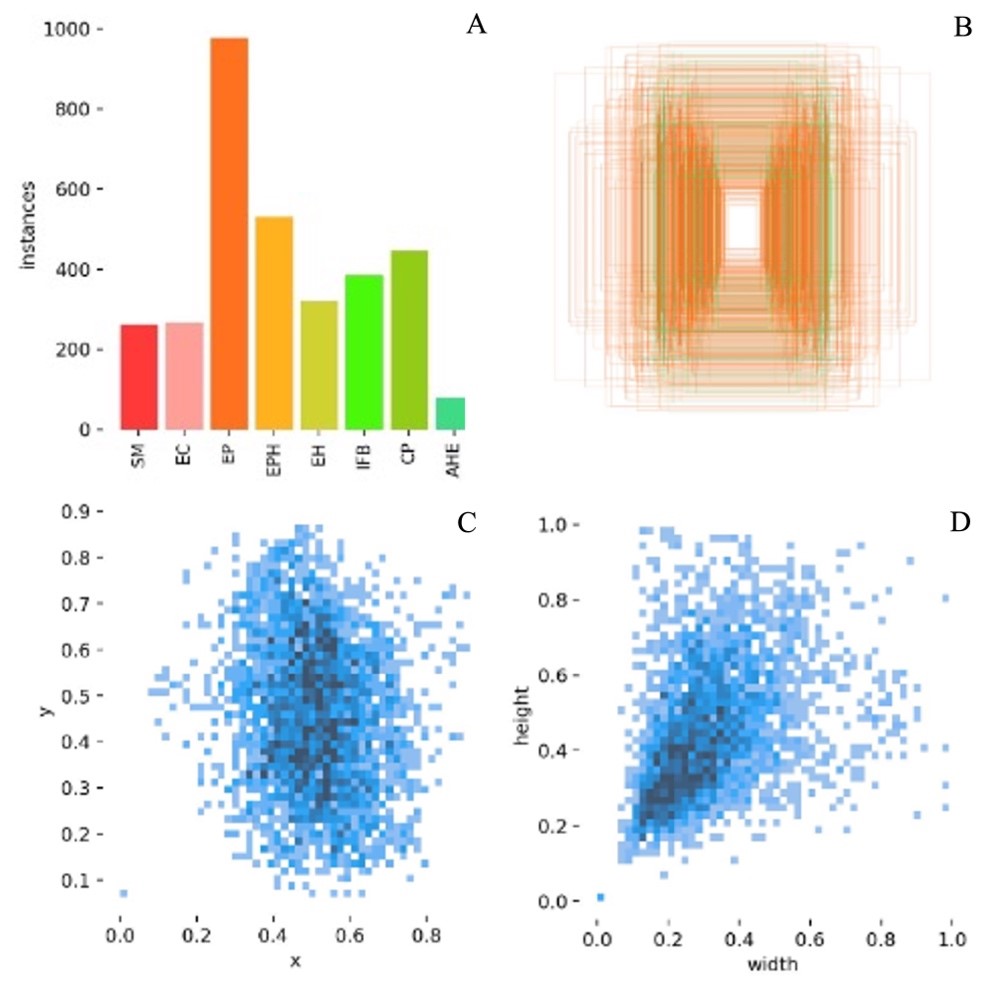}
    \caption{\label{f1}Detail and characteristic of the dataset. Statistics of different categories and the position and size of the bounding box.\newline
A. Statistics of different labels; B. different labels bounding box size, the color corresponds to the label color in A; C. map of bounding box location distribution; D. map of bounding box size distribution.\newline
SM, submucous myoma; EC, endometrial cancer; EP, endometrial polyp; EPH, endometrial polypoid hyperplasia; EH, endometrial hyperplasia without atypical hyperplasia; IFB, intrauterine foreign body; CP, cervical polyp; AHE, atypical hyperplasia of endometrium.
}
    \label{fig:enter-label}
\end{figure}

\section{Experimental Design, Materials and Methods}
A total of 175 videos from 175 patients were collected from Beijing Chaoyang Hospital, Capital Medical University from August 2023 to January 2024. All included patients had tissue pathology results confirming their diagnosis. The study was approved by the Medical Ethics Committee of Beijing Chaoyang Hospital Affiliated to Capital Medical University (2023-S-454). All authors confirm that we have complied with all relevant ethical regulations. All images were taken using Karl Storz TC200 endoscopic camera systems with a resolution of 1920 × 1080 pixels and were stored in MP4 format. Images are extracted from videos in 10 frames each. Images meeting the specified criteria were excluded for the following reasons: (a) poor quality or unclear content; (b) absence of lesions within the field of view; (c) presence of substantial bleeding in the field of view. All images were marked with lesion sites by gynecologists. We used Yolov5l6 model and set 8 classes to train our dataset. Finally, a total of 3385 images of different types of endometrial lesions were collected. We randomly extracted 686 images from the dataset as the validation set and the remaining 2745 images were used as the original training set for data augmentation and model training, setting net input size as 1280 pixel.

\begin{figure}
    \centering
    \includegraphics[width=\textwidth]{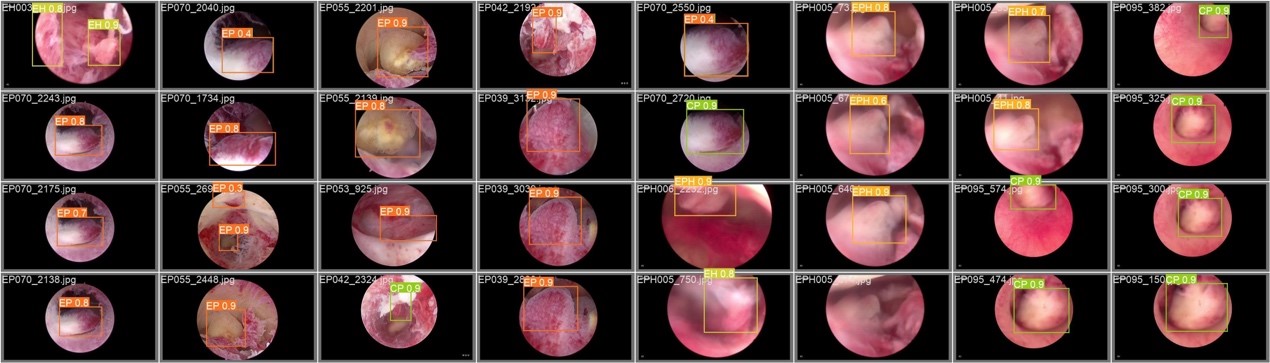}
    \caption{The bounding boxes manipulation for the lesion tissue detection by Yolov5l6 model.}
    \label{fig:enter-label}
\end{figure}

\section{Result}
We evaluate models’ performances using a combination of Recall (R) and Precision (P) scores with various Intersection over Union (IoU) thresholds, which included mAP50 and mAP [0.5,0.95].

\begin{table}
\centering
\caption{\label{t2}Models’ performances in diagnosing different endometrial diseases.}
\begin{threeparttable} 
\begin{tblr}{
  width = \linewidth,
  colspec = {Q[87]Q[167]Q[169]Q[227]Q[279]},
  cells = {c},
  cells = {m},
  hline{1,11} = {-}{0.08em},
  hline{2} = {-}{},
}
class & P         & R         & AP50         & AP50-95         \\
all   & 0.824(mP) & 0.756(mR) & 0.799(mAP50) & 0.453(mAP50-95) \\
SM    & 0.841     & 0.923     & 0.947        & 0.693           \\
EC    & 0.834     & 0.716     & 0.741        & 0.311           \\
EP    & 0.865     & 0.796     & 0.874        & 0.526           \\
EPH   & 0.898     & 0.693     & 0.759        & 0.382           \\
EH    & 0.648     & 0.492     & 0.522        & 0.21            \\
IFB   & 0.885     & 0.838     & 0.868        & 0.487           \\
CP    & 0.83      & 0.771     & 0.854        & 0.538           \\
AHE   & 0.787     & 0.823     & 0.823        & 0.476           
\end{tblr}
\begin{tablenotes} 
\item[*] P, precision; R, recall; AP, average precision; mP, mean precision; mR, mean recall; mAP, mean average precision.
SM, submucous myoma; EC, endometrial cancer; EP, endometrial polyp; EPH, endometrial polypoid hyperplasia; EH, endometrial hyperplasia without atypical hyperplasia; IFB, intrauterine foreign body; CP, cervical polyp; AHE, atypical hyperplasia of endometrium.
\end{tablenotes}
\end{threeparttable}
\end{table}

\begin{figure}
    \centering
    \includegraphics[width=\textwidth]{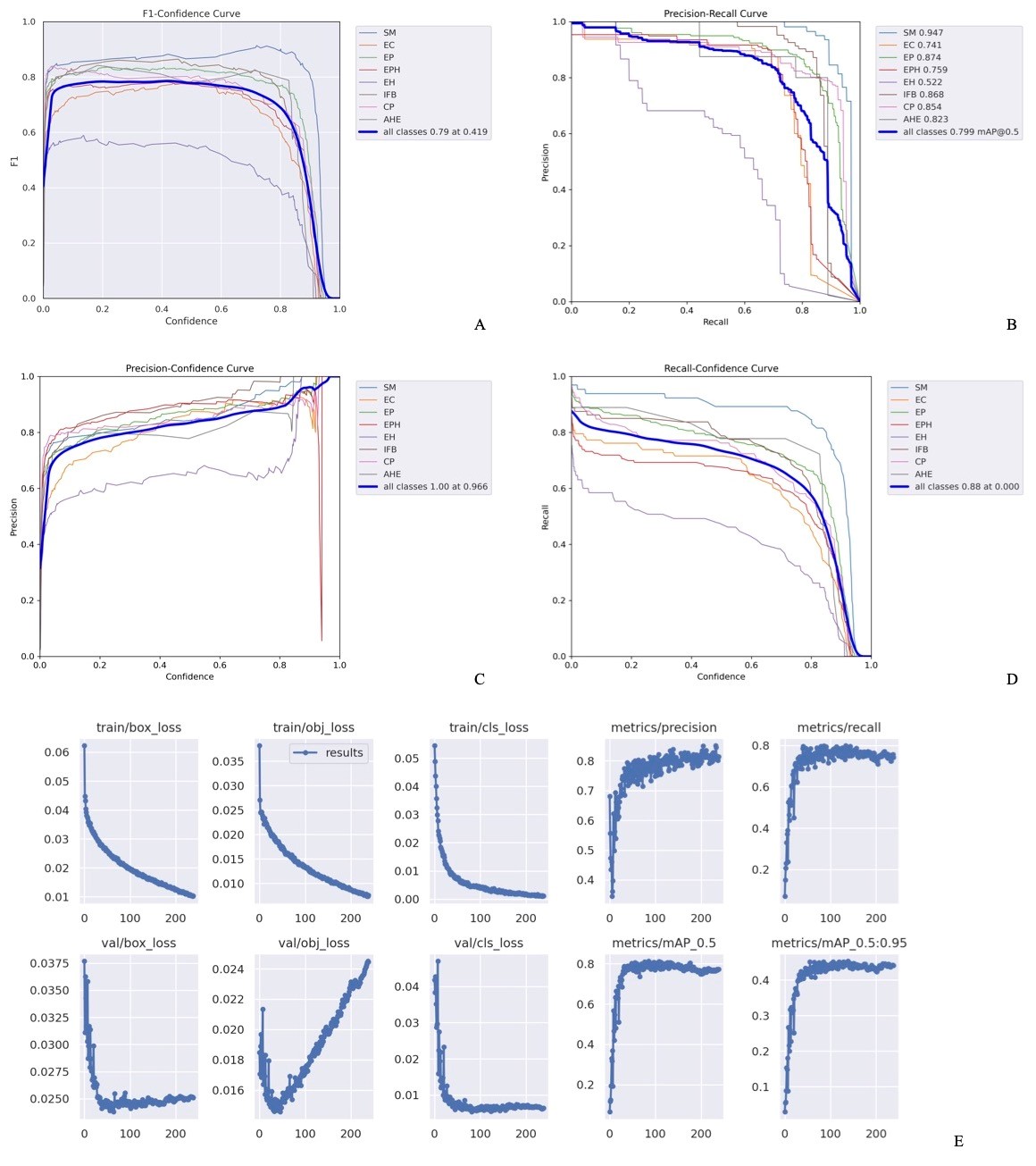}
    \caption{Result of train. A. F1-confidence curve for all labels. B. precision-recall curve. C. precision-confidence curve. D. recall-confidence curve. E. train loss and val loss. detail, box\_loss is regression loss of bounding box, obj\_loss is object detect loss, cls\_loss is classification loss. F1, the harmonic mean of the precision and recall. Precision, recall, mAP\_0.5, mAP\_0.5:0.95 is all classes mean curve. 
}
    \label{fig:enter-label}
\end{figure}

\section{Conclusion}
This article proposes a uterine cavity dataset, which includes hysteroscopic images of various gynecological benign and malignant diseases and labels confirmed by doctors with clinical diagnostic results. Given that there is currently no dataset available for training deep learning algorithms related to uterine cavity lesions. We propose this dataset and use the YOLO algorithm to train the detection model as the baseline algorithm based on this data. We hope that this dataset can serve as the data basis for developing an algorithm training software for assisting in the diagnosis of benign and malignant intrauterine diseases based on hysteroscopic data, thereby achieving assisted diagnosis in hysteroscopic surgery.

\section*{Ethics Statements}
The study was approved by the Medical Ethics Committee of Beijing Chao-Yang Hospital, Capital Medical University (2023-S-454).
Declaration of Competing Interest
The authors declare that they have no known competing financial interests or personal relationships that could have appeared to influence the work reported in this paper.

\bibliographystyle{ieeetr}

\end{CJK}
\end{document}